\documentclass[aps,twocolumn,showpacs,pra]{revtex4}
\usepackage{amsmath}
\usepackage{amsfonts}
\usepackage{amssymb}
\usepackage{epsfig}
\usepackage{graphicx}
\begin{document}
\title{Electromagnetically induced transparency and optical memories in an optomechanical system with $N$ membranes}
\author{Sumei Huang$^{1}$ and Mankei Tsang$^{1,2}$}

\address{$^{1}$ Department of Electrical and Computer Engineering,\\
National University of Singapore, 4 Engineering Drive 3, Singapore 117583}
\address{$^{2}$ Department of Physics, National University of Singapore, 2 Science Drive
3, Singapore 117551}

\date{\today}
\begin{abstract}
We study the propagation of a weak probe field through an optomechanical system in which $N$ nearly degenerate mechanical membranes are inside a Febry-Perot cavity, and couple dispersively to an intracavity field. We derive a general analytical expression for the output probe field. We show that the in-phase quadrature of the output probe field can exhibit multiple electromagnetically induced transparency with at most $N$ narrow transparency windows. We find that the superluminal light can be achieved with at most $N$ different negative group velocities at the transparency windows. Moreover, by using a time-dependent strong coupling pulse, we numerically simulate the storage and retrieval of a weak Gaussian-shaped probe pulse in an optical cavity with two nearly degenerate mechanical resonators.
\end{abstract}
\maketitle
\section{Introduction}

Electromagnetically induced transparency (EIT) is a quantum destructive interference effect that allows the propagation of a weak coherent probe field through a three-level atomic medium in the $\Lambda$-configuration in the presence of a strong coupling field \cite{Harris1,Harris2}. It is found that the transmitted light can propagate with group velocities smaller than the speed of light in vacuum $c$ (slow light) \cite{Hau,Kash}, or greater than $c$ or even with negative group velocities (superluminal light) \cite{fast1,fast2,Zhu}. The superluminal light with a negative group velocity $\sim-c/14400$ has been observed in a Cs atomic vapor system \cite{fast2}. Moreover, the studies of the EIT have been extended to multilevel atomic systems interacting with multiple laser beams \cite{Knight1,Knight2,Xiao2,Lukin,Gavra,Goren}, whose absorption profiles of a weak probe field are different from that of three-level $\Lambda$ system. It has been shown that multiple EIT windows in the probe absorption spectrum are observed in the multilevel atomic systems \cite{Knight1}. Multiple EIT windows allow transmissions of the probe light at multiple different frequencies simultaneously. Such multiple EIT could be useful for multi-channel optical communication and multi-channel quantum information processing. In addition, it has been demonstrated that light pulses can be stored in atomic coherences with long lifetime by using the EIT in atomic mediums \cite{Dutton, Mair,Heinze}. The EIT-based light storage is a promising technique, which leads to the realization of
optical memories \cite{Dutton, Mair,Heinze}.

Recently, studies of the EIT effect have been extended to the macroscopic optomechanical systems. It has been pointed out that the existence of an analogy of the EIT in the optomechanical systems \cite{Agarwal1}. And some experiments have been reported for observing the EIT-like dips in the optomechanical systems \cite{Weis, Lin, Safavi, Teufel}. Besides, the EIT in membrane-in-the-middle setup at room temperature has been demonstrated experimentally \cite{Vitali}. Additionally, the EIT in quadratically coupling optomechanical systems and the EIT in optomechanical systems using quantized fields have been analyzed \cite{Sumei1,Sumei2}. Quite recently, the EIT in the nonlinearized optomechanical systems has been discussed theoretically \cite{Lemonde,Girvin,Kronwald}. In addition, it has been shown experimentally and theoretically that optomechanical systems can be used as optical memory elements \cite{Hailin,Qug} based on the EIT effect. On the other hand, optomechanical systems with multi-membranes coupled to a common cavity mode via radiation pressure have been studied, including cooling and trapping of mechanical modes \cite{Meystre2008,Meystre20082}, quantum state transfer and entanglement between mechanical modes \cite{Meystre2012, Jack, Meystre, Plenio, Liu, Schmidt}.

In this paper, we investigate the response of a Fabry-Perot cavity with $N$ membranes having close frequencies to a weak probe field in the absence and the presence of a strong coupling field. We find that there are at most $N$ transparency windows in the transmitted probe field in the presence of the coupling field. We also show that the group velocity can obtain $N$ different values at the transparency and can be manipulated by changing the effective optomechanical coupling strengths. Additionally, we present that an optical cavity with two membranes can store and retrieve a probe pulse at two different frequencies.

The paper is organized as follows. In Sec. II, we introduce the system with $N$ membranes interacting with a single cavity mode, give the equations of motion for the mean value of the system operators, and obtain the general analytical expression for the component of the output field at the probe frequency. In Sec. III, we present the numerical results for the output probe field from an optomechanical system with several membranes. In Sec. IV, we demonstrate that the probe pulse at two different frequencies can be stored in mechanical excitations of the two membranes by applying a writing coupling pulse and can
be retrieved later by applying a reading coupling pulse. Finally, we conclude in Sec. V.

\section{Model}
\begin{figure}[!h]
\begin{center}
\scalebox{0.45}{\includegraphics{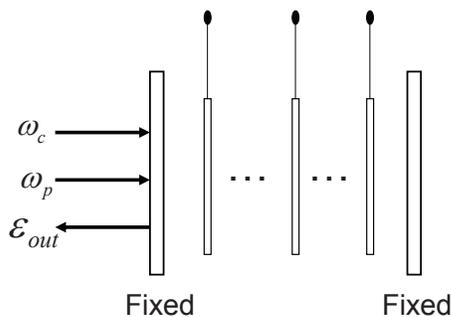}} \caption{\label{Fig1} Sketch of the studied system.  A strong coupling field at
frequency $\omega_{c}$ and a weak probe field at frequency $\omega_{p}$ are sent into
 an optical cavity consisting of two rigidly fixed mirrors. One mirror is partially transparent and the other one is totally reflecting. The $N$ nonabsorbing membranes
are placed inside the cavity. After the interaction
between the cavity field and the membranes, the output field $\varepsilon_{out}$ will
contain three frequencies ($\omega_{c}$, $\omega_{p}$, and $2\omega_{c}-\omega_{p}$).}
\end{center}
\end{figure}

 We consider an optical Fabry-Perot cavity containing $N$ lossless membranes with effective mass $m_{n}$, resonance frequency $\omega_n$, and damping rate $\gamma_{n}$ $(n = 1, . . . , N)$, as shown in Fig.~\ref{Fig1}. The thickness of each membrane is much smaller than the wavelength of the cavity field, and each membrane is partially transparent. The cavity field is driven by a strong coupling field at frequency $\omega_{c}$. Meanwhile a weak probe field at frequency $\omega_{p}$ is injected into the cavity. The intracavity photons exert radiation pressure forces on each membrane so that each membrane makes small oscillations. As a result, the resonance frequency of the cavity field depends on the displacement $q_{n}$ of each membrane, represented by $\omega_{0}(\{q_{n}\})$. Here we consider the case that all the membranes are located at the nodes of the frequency $\omega_{0}(\{q_{n}\})$ of the cavity field.
Thus the cavity frequency $\omega_{0}(\{q_{n}\})$ depends linearly on $q_{n}$,
 \begin{equation}\label{1}
 \omega_{0}(\{q_{n}\})=\omega_{0}+\sum_{n=1}^{N}g_{n0}q_{n},
 \end{equation}
 where $\omega_{0}$ is the cavity resonance frequency for $q_{n}=0$, $g_{n0}=\frac{\partial \omega_{0}(\{q_{n}\})}{\partial q_{n}}|_{q_{n}=0}$ is the optomechanical coupling strength between the cavity mode and the $nth$ membrane.

 For convenience, we write the position and momentum operators $(q_{n}, p_{n})$ of the membranes in terms of the dimensionless variables $(Q_{n}, P_{n})$ with $q_{n}=\sqrt{\frac{\hbar}{m_{n}\omega_{n}}}Q_{n}$, $p_{n}=\sqrt{m_{n}\hbar\omega_{n}}P_{n}$, and $[Q_{j}, P_{k}]=i\delta_{jk}$. In a frame rotating at the driving frequency $\omega_{c}$, the Hamiltonian of the whole system reads
 \begin{eqnarray}\label{2}
H&=&\hbar(\omega_{0}-\omega_{c})c^{\dag}c+\sum_{n=1}^N \frac{\hbar\omega_{n}}{2}(Q_{n}^2+P_{n}^2)\nonumber\\
& &+\sum_{n=1}^N\hbar g_{n}Q_{n}c^{\dag}c+i\hbar \varepsilon_{L} (c^{\dag}-c)\nonumber\\
& &+i\hbar(\varepsilon_{p}c^{\dag}e^{-i\delta t}-\varepsilon^{*}_{p}ce^{i\delta t}).
\end{eqnarray}
 Here the first two terms are the energies of the cavity field and the $N$ mechanical oscillators, respectively. The $c^{\dag}$ and $c$ are the creation and annihilation operators of the cavity field, obey the standard commutation relation $[c,c^{\dag}]=1$, $c^{\dag}c$ is the photon number operator of the cavity field. The third term gives the optomechanical interactions between the cavity field and the membranes, $g_{n}=g_{n0}\sqrt{\frac{\hbar}{m_{n}\omega_{n}}}$. The last two terms describe the interactions of the cavity field with the coupling field and the probe field, respectively. The $\varepsilon_{L}$ quantifies the coupling strength between the coupling field and the cavity field, depends on the power $\wp$ of the coupling field by $\varepsilon_{L}=\sqrt{\frac{2\kappa \wp}{\hbar \omega_{c}}}$, where $\kappa$ is the photon loss rate of the cavity through the fixed mirror. The $\varepsilon_{p}$ represents the coupling strength between the probe field and the cavity field, it is related to the power $\wp_{p}$ of the probe field by $\varepsilon_{p}=\sqrt{\frac{2\kappa \wp_{p}}{\hbar \omega_{p}}}$. The $\delta=\omega_{p}-\omega_{c}$ is the detuning of the probe field from the coupling field. In the following, we are interested in the regime that the mechanical damping rate is much smaller than the photon decay rate $\gamma_{n}\ll\kappa$.

Taking account of the effects of dissipations of the cavity field and the mechanical oscillators, and neglecting quantum noise and thermal noise, we obtain the time evolutions of the expectation values of the system operators
\begin{eqnarray}\label{3}
\langle \dot{Q}_{n}\rangle&=&\omega_{n}\langle P_{n}\rangle,\nonumber\\
\langle \dot{P}_{n}\rangle&=&-\omega_{n}\langle Q_{n}\rangle-g_{n}\langle c^{\dag}\rangle \langle c\rangle-\gamma_{n}\langle P_{n}\rangle,\nonumber\\
\langle \dot{c}\rangle&=&-\big\{\kappa+i\big[\omega_{0}-\omega_{c}+\sum_{n=1}^{N}g_{n}\langle Q_{n}\rangle\big]\big\}\langle c\rangle+\varepsilon_{L}\nonumber\\& &+\varepsilon_{p}e^{-i\delta t}.
\end{eqnarray}
Here we focus on the strong-driving regime, so the mean field assumption $\langle c^{\dag}c\rangle\simeq\langle c^{\dag}\rangle\langle c\rangle$ and $\langle Q_{n}c\rangle\simeq\langle Q_{n}\rangle\langle c\rangle$ has been used in Eq. (\ref{3}). Since the probe field is much weaker than the coupling field $(|\varepsilon_{p}|<<\varepsilon_{L})$,
the steady-state solution to Eq. (\ref{3}) can be approximated to the first order in the probe field $\varepsilon_{p}$. In the long time limit, the solution to Eq. (\ref{3}) can be written as
\begin{eqnarray}\label{4}
\langle s\rangle=s_{0}+s_{+}\varepsilon_{p}e^{-i\delta t}+s_{-}\varepsilon_{p}^{*}e^{i\delta t},
\end{eqnarray}
where $s=Q_{n}$, $P_{n}$, or $c$. The solution contains three components, which in the original
frame oscillate at $\omega_{c}$, $\omega_{p}$, $2\omega_{c}-\omega_{p}$, respectively. Substituting Eq. (\ref{4}) into Eq. (\ref{3}), equating coefficients of $e^{0}$ and $e^{\pm i\delta t}$, we can obtain $c_{0}$, $c_{+}$, and $c_{-}$.

The output field can be obtained by using the input-output relation $\varepsilon_{out}=2\kappa \langle c\rangle$ \cite{Walls}.
In analogy with Eq. (\ref{4}), we expand the output field to the first order in the probe field $\varepsilon_{p}$,
\begin{eqnarray}\label{5}
\varepsilon_{out}=\varepsilon_{out0}+\varepsilon_{out+}\varepsilon_{p}e^{-i\delta t}+\varepsilon_{out-}\varepsilon_{p}^{*}e^{i\delta t},
\end{eqnarray}
where $\varepsilon_{out0}$, $\varepsilon_{out+}$, and $\varepsilon_{out-}$ are the components of the output field oscillating at frequencies $\omega_{c}$, $\omega_{p}$, $2\omega_{c}-\omega_{p}$. Hence, the component of the output field at the probe frequency $\omega_{p}$ which we are interested in is $\varepsilon_{out+}=2\kappa c_{+}$.

We assume the frequencies of the $N$ membranes are slightly different, and the mean value of them is $\omega_{m}$. If $\delta=\Delta$, and $\delta=\omega_{m}$, the interactions between the cavity field and the membranes are nearly the strongest. In addition, it is assumed that the frequencies of the $N$ membranes are much larger than the cavity decay rate $\omega_{n}\gg \kappa$. Under these conditions, the component of the output field at the probe frequency is
\begin{equation}\label{6}
\varepsilon_{out+}\simeq\frac{2\kappa}{\kappa-i(\delta-\omega_{m})+\sum_{n=1}^{N}\frac{G_{n}^2/2}{\frac{\gamma_n}{2}-i(\delta-\omega_{n})}},
\end{equation}
where $G_{n}=g_{n}|c_{0}|$ is the effective optomechanical coupling rate, $c_{0}=\frac{\varepsilon_{L}}{\kappa+i\Delta}$, $\Delta=\omega_{0}-\omega_{c}+\sum_{n=1}^{N}g_{n}Q_{n0}$ is the effective detuning of the coupling field from the cavity resonance frequency, including the frequency shift induced by the radiation pressure. Let us write $\varepsilon_{out+}$ as $\varepsilon_{out+}=\upsilon_{p}+i\tilde{\upsilon}_{p}$, where $\upsilon_{p}$ and $\tilde{\upsilon}_{p}$ are the in-phase and out-of-phase quadratures of the output probe field, representing the absorptive and dispersive properties of the output probe field, respectively. The quadratures can be measured via the homodyne technique \cite{Walls}.

  When $\gamma_{n}\ll\kappa$, $\varepsilon_{out+}$ is extremely small at $\delta=\omega_{n}$ ($n=1,2,3,...,N$) . Thus if all the frequencies of the $N$ membranes are different, the optomechanical system becomes transparent at $N$ different frequencies of the probe field in the weak coupling regime $\sqrt{2}G_{n}<\kappa$. The full linewidth at half maximum (FWHM) of each EIT dip is about $\gamma_{n}+\frac{G_{n}^2}{\kappa}$, which can be broadened by increasing the power of the coupling field.

 Next we assume that $L$ ($1<L<N$) of the frequencies of the membranes are equal to $\omega$, i.e., ($\omega_{1}=\omega_{2}=...=\omega$) and the left $N-L$ are different from $\omega$, the output field at the probe frequency becomes
 \begin{widetext}
 \begin{equation}\label{7}
\varepsilon_{out+}\simeq\frac{2\kappa}{\kappa-i(\delta-\omega_{m})+\sum_{n=1}^{L}\frac{G_{n}^2/2}{\frac{\gamma_n}{2}-i(\delta-\omega)}+\sum_{n=L+1}^{N}\frac{G_{n}^2/2}{\frac{\gamma_n}{2}-i(\delta-\omega_{n})}}.
\end{equation}
\end{widetext}
If $\sqrt{2}G_{n}<\kappa$, $N-L+1$ transparency windows exist in the output probe field.

Finally if all the $N$ membranes have the same frequency $\omega_{n}=\omega_{m}$,
 the same damping rate $\gamma_{n}=\gamma$, and the same coupling rate $G_{n}=G$, the output probe field reduces to
\begin{equation}\label{8}
\varepsilon_{out+}\simeq\frac{2\kappa}{\kappa-i(\delta-\omega_{m})+\frac{NG^2/2}{\frac{\gamma}{2}-i(\delta-\omega_{m})}},
\end{equation}
whose form is similar to that of an optomechanical system with a single mechanical oscillator \cite{Agarwal1}.
If $\sqrt{2N}G<\kappa$, there is only one transparency window at line center $\delta=\omega_{m}$ in the output probe field, its FWHM is about $\gamma+\frac{NG^2}{\kappa}$.

On the other hand, the group velocity of the output probe field can be determined by \cite{HarrisF}
\begin{equation}\label{9}
v_{g}=\frac{c}{1+\frac{1}{2}\tilde{\upsilon}_{p}+\frac{\delta}{2}\frac{\partial \tilde{\upsilon}_{p}}{\partial\delta}},
\end{equation}
where $c$ is the velocity of light in vacuum. It is noticed that the group velocity $v_{g}$ depends on the out-of-phase quadrature $\tilde{\upsilon}_{p}$, the slope $\frac{\partial \tilde{\upsilon}_{p}}{\partial\delta}$, and the probe detuning $\delta$.

\section{Numerical results of the output probe field}
In this section, we numerically evaluate the quadratures $\upsilon_{p}$, $\tilde{\upsilon}_{p}$, and the group velocity $v_{g}$ of the output probe field when the cavity have $N$ membranes whose frequencies are very close.

The values of the parameters chosen are similar to those in \cite{Aspelmeyer}: the mean value of the frequencies of the $N$ membranes is $\omega_{m}=2\pi\times134$ kHz, the cavity decay rate $\kappa=\omega_{m}/5$, ($\kappa/\omega_{m}=0.2<1$, the system is in the resolved sideband regime), the mechanical damping rates are taken to be equal $\gamma_{n}=2\pi\times 0.12$ Hz.

\begin{figure}[!h]
\begin{center}
\scalebox{0.6}{\includegraphics{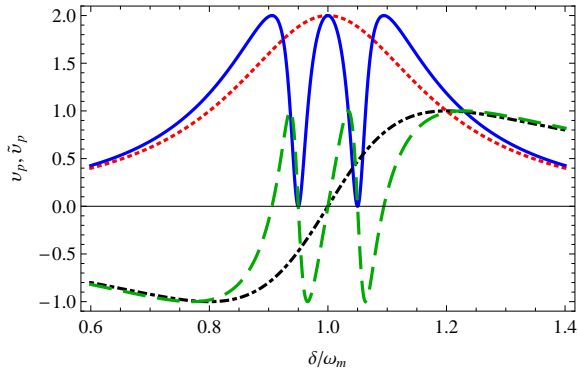}} \caption{\label{Fig2} (Color online) The quadratures $\upsilon_{p}$ (dotted and solid curves) and $\tilde{\upsilon}_{p}$ (dot-dashed and dashed curves) of the output probe field as a function of $\delta/\omega_{m}$ in the absence (dotted and dot-dashed curves) and the presence (solid and dashed curves) of the coupling field for a system with two membranes. Parameters: $\omega_{1}=\omega_{m}+0.05\omega_{m}$, $\omega_{2}=\omega_{m}-0.05\omega_{m}$, $G_{1}=G_{2}=0.4\kappa$.}
\end{center}
\end{figure}

\begin{figure}[!h]
\begin{center}
\scalebox{0.6}{\includegraphics{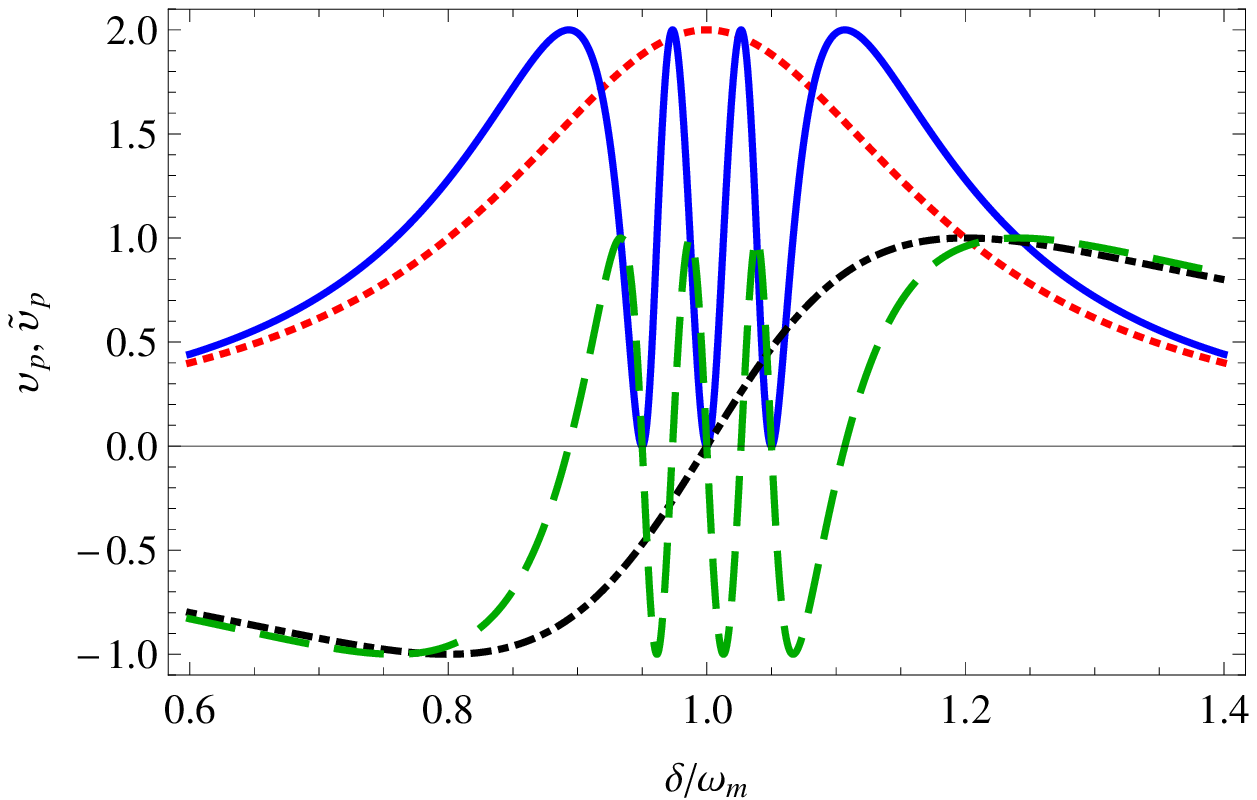}} \caption{\label{Fig3} (Color online) Same as in Fig. \ref{Fig2} except for a system with three membranes. Parameters: $\omega_{1}=\omega_{m}+0.05\omega_{m}$, $\omega_{2}=\omega_{m}$, $\omega_{3}=\omega_{m}-0.05\omega_{m}$, $G_{1}=G_{2}=G_{3}=0.4\kappa$.}
\end{center}
\end{figure}

\begin{figure}[!h]
\begin{center}
\scalebox{0.6}{\includegraphics{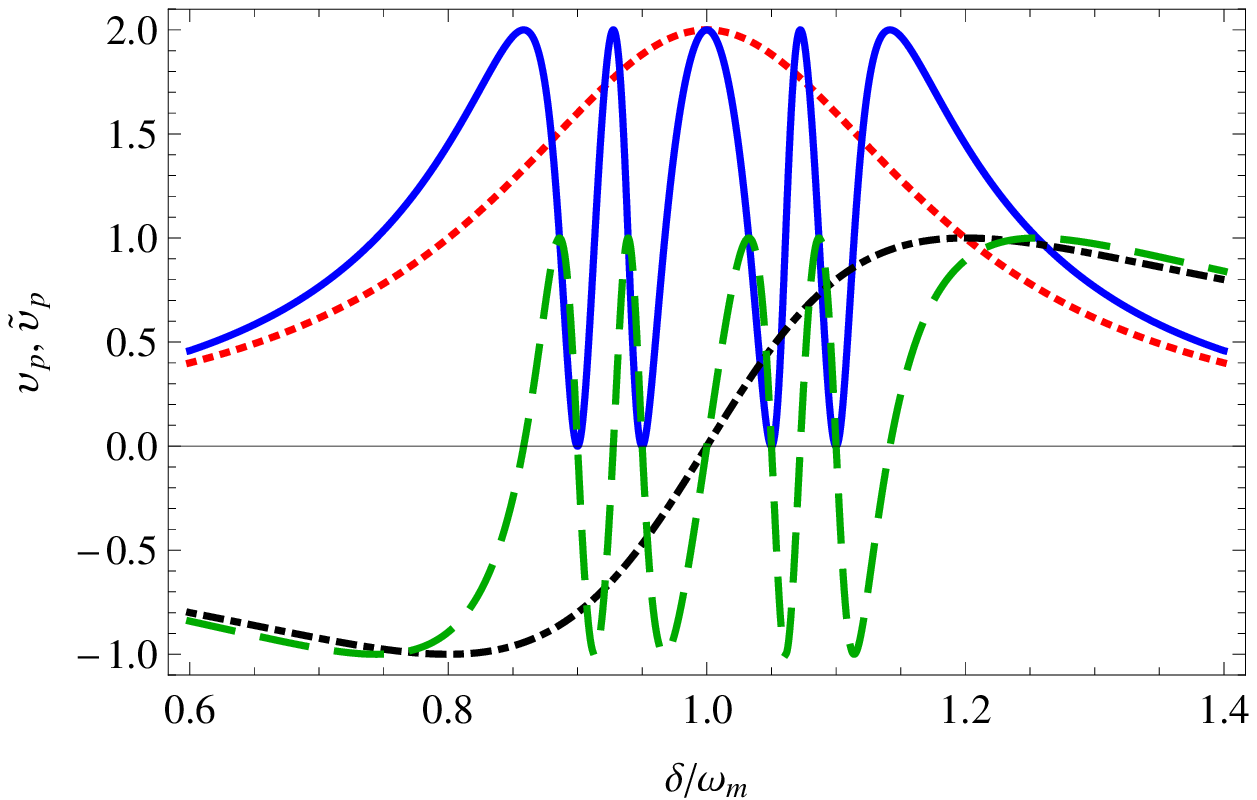}} \caption{\label{Fig4} (Color online) Same as in Fig. \ref{Fig2} except for a system with four membranes. Parameters: $\omega_{1}=\omega_{m}+0.05\omega_{m}$, $\omega_{2}=\omega_{m}-0.05\omega_{m}$, $\omega_{3}=\omega_{m}+0.1\omega_{m}$, $\omega_{4}=\omega_{m}-0.1\omega_{m}$, $G_{1}=G_{2}=G_{3}=G_{4}=0.4\kappa$.}
\end{center}
\end{figure}
\begin{figure}[!h]
\begin{center}
\scalebox{0.6}{\includegraphics{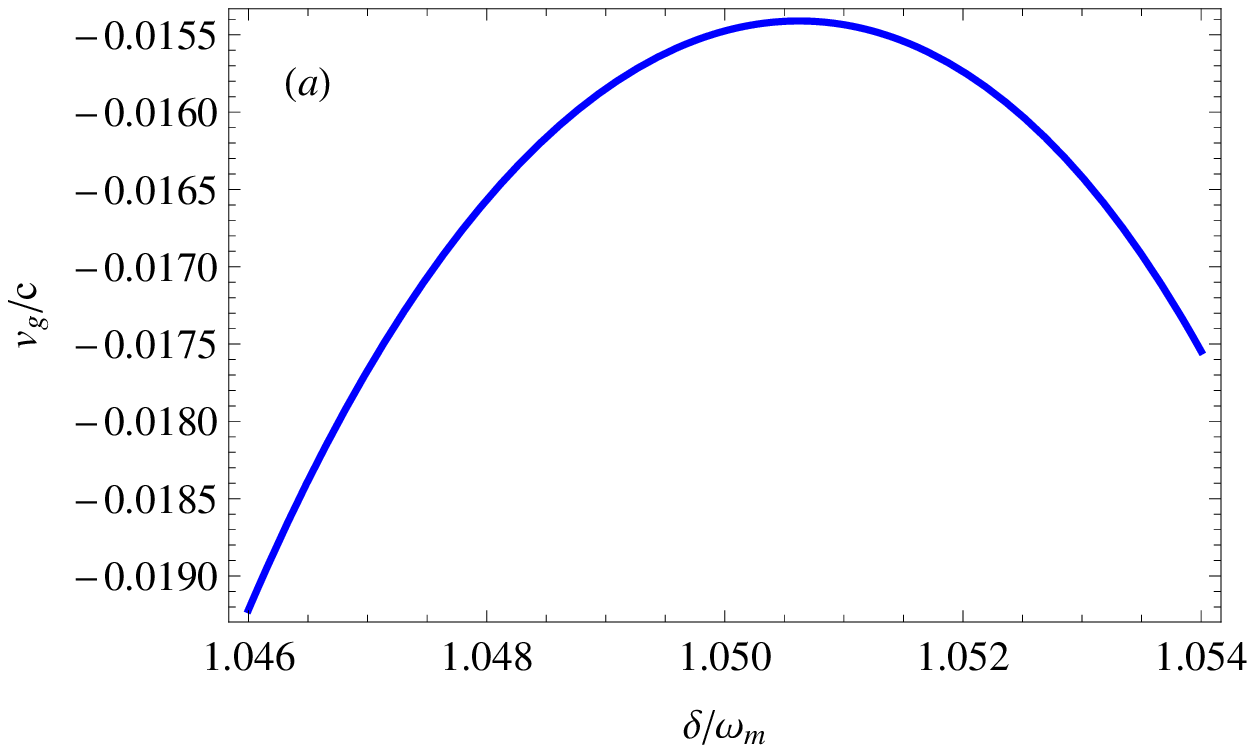}} \\
\scalebox{0.6}{\includegraphics{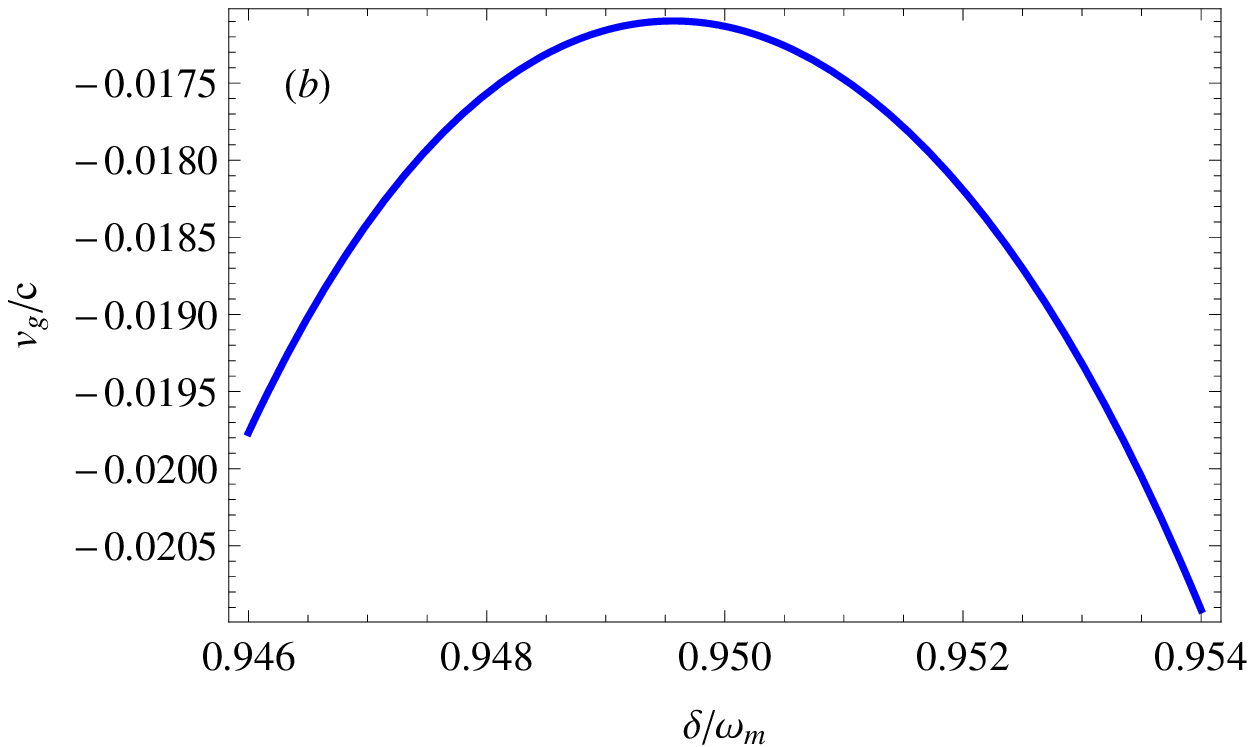}} \\
\caption{\label{Fig5} (Color online) The normalized group velocities $v_{g}/c$ at the two transparency windows in Fig.~\ref{Fig2} as a function of $\delta/\omega_{m}$. Parameters: $\omega_{1}=\omega_{m}+0.05\omega_{m}$, $\omega_{2}=\omega_{m}-0.05\omega_{m}$, $G_{1}=G_{2}=0.4\kappa$.}
\end{center}
\end{figure}

 We plot the phase quadratures $\upsilon_{p}$ and $\tilde{\upsilon}_{p}$ of the output probe field as a function of the normalized probe detuning $\delta/\omega_{m}$ for a system with two membranes (Fig.~\ref{Fig2}), three membranes (Fig.~\ref{Fig3}), and four membranes (Fig.~\ref{Fig4}). We assume that all membranes are coupled to the cavity field with the same optomechanical coupling strength $G_{n}=0.4\kappa$. In the absence of the coupling field, $\upsilon_{p}$ (dotted curve) has the standard Lorentzian absorption shape, whereas $\tilde{\upsilon}_{p}$ (dot-dashed curve) has a standard dispersion shape. In the presence of the coupling field, $\upsilon_{p}$ (solid curve) exhibits narrow two, three, and four transparency windows in Figs.~\ref{Fig2}-\ref{Fig4}. The multiple transparency windows display that the optomechanical system becomes simultaneously transparent to the probe field at multiple different frequencies, which is the result of the destructive interferences between the input probe field and the anti-Stokes fields generated by the interactions of the coupling field with the multi-membranes. Moreover, We also note that the EIT dips are accompanied by steep variations of $\tilde{\upsilon}_{p}$ (dashed curve) with identical negative slopes, which lead to the generation of superluminal light with nearly equal group velocities. For a cavity with two membranes, the group velocities $v_{g}$ around the two transparency windows are shown in Fig.~\ref{Fig5}. At the transparency points $\delta/\omega_{m}=1.05, 0.95$, the group velocities are $v_{g}=-0.0155c, -0.0171c$, respectively.

\begin{figure}[!h]
\begin{center}
\scalebox{0.6}{\includegraphics{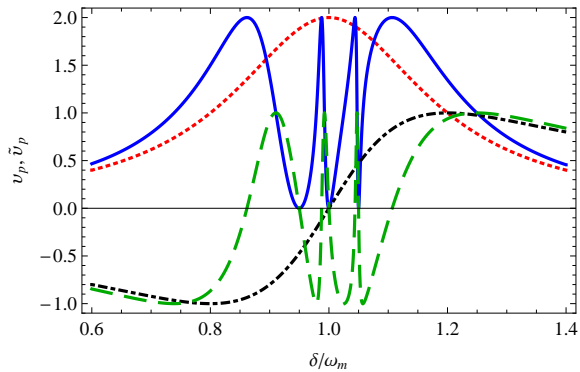}} \caption{\label{Fig6} (Color online) Same as in Fig. \ref{Fig2} except  for a system with three membranes. Parameters:  $\omega_{1}=\omega_{m}+0.05\omega_{m}$, $\omega_{2}=\omega_{m}$, $\omega_{3}=\omega_{m}-0.05\omega_{m}$, $G_{1}=0.2\kappa$, $G_{2}=0.4\kappa$, $G_{3}=0.7\kappa$.}
\end{center}
\end{figure}
We also plot the phase quadratures $\upsilon_{p}$ and $\tilde{\upsilon}_{p}$ of the output probe field for a system with three membranes (Fig.~\ref{Fig6}). But each membrane is coupled to the optical mode with nonidentical optomechanical coupling strengths $G_{1}=0.2\kappa$, $G_{2}=0.4\kappa$, $G_{3}=0.7\kappa$. It is seen that three transparency windows with different linewidths appear in the quadrature $\upsilon_{p}$ of the output probe field. And the negative slopes of $\tilde{\upsilon}_{p}$ at the transparency are different, which results in three different group velocities of the output probe light at three different frequencies $\delta/\omega_{m}=1.05$, $1$, $0.95$. Our calculations show $v_{g}=-0.0038c$ at $\delta/\omega_{m}=1.05$, $v_{g}=-0.0163c$ at $\delta/\omega_{m}=1$, $v_{g}=-0.0544c$ at $\delta/\omega_{m}=0.95$.

\begin{figure}[!h]
\begin{center}
\scalebox{0.6}{\includegraphics{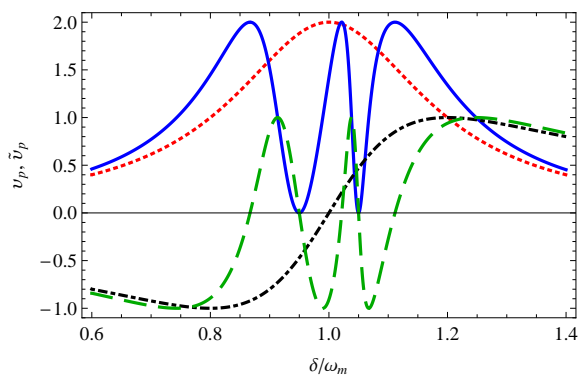}} \caption{\label{Fig7} (Color online) Same as in Fig. \ref{Fig2} except for a system with four membranes. Parameters: $\omega_{1}=\omega_{m}+0.05\omega_{m}$, $\omega_{2}=\omega_{3}=\omega_{4}=\omega_{m}-0.05\omega_{m}$, $G_{1}=G_{2}=G_{3}=G_{4}=0.4\kappa$.}
\end{center}
\end{figure}
Finally, in Fig.~\ref{Fig7}, we give an example about a system with four membranes, but three of them having identical frequency, i.e., $\omega_{1}=\omega_{m}+0.05\omega_{m}$, $\omega_{2}=\omega_{3}=\omega_{4}=\omega_{m}-0.05\omega_{m}$. Note that only two transparency windows with different linewidths occur in the quadrature $\upsilon_{p}$ of the output probe field. The transmitted probe light at the two different frequencies $\delta/\omega_{m}=1.05,0.95$ are propagating at the different group velocities $v_{g}=-0.015c,-0.053c$.

\section{Storage and Retrieval of light in a cavity with two membranes}
 Since the mechanical membrane damping rate $\gamma_n$ is much smaller than the cavity decay rate $\kappa$, an optical cavity with multiple membranes can store a weak probe field at different frequencies as long-lived mechanical excitations in the membranes simultaneously under the EIT condition. Here we take a cavity with two membranes
for example. We show the possibility of storing and retrieving a probe light pulse at two different frequencies in such a system. We assume that the probe laser pulse and the coupling laser pulse are time-dependent, and they have Gaussian shapes
\begin{eqnarray}
\varepsilon_{p}(t)&=&\varepsilon_{p}\exp{\big[-\frac{(t-t_{wr})^2}{2\tau_{p}^2}\big]},\nonumber\\
\varepsilon_{L}(t)&=&\varepsilon_{L}\exp{\big[-\frac{(t-t_{wr})^2}{2\tau_{L}^2}\big]}+\varepsilon_{L}\exp{\big[-\frac{(t-t_{rd})^2}{2\tau_{L}^2}\big]},\nonumber\\
\end{eqnarray}
where $t_{wr}$ and $t_{rd}$ are the central time of the writing and reading coupling lasers. The difference $t_{rd}-t_{wr}$ gives
the storage time. The $\tau_{p}$ and $\tau_{L}$ are the widths of the probe pulse and the coupling pulse, respectively. We assume that $\tau_{p}\leq\tau_{L}$ and $\tau_{p}^{-1}$ is less than the FWHM of EIT dip $(\gamma_{n}+\frac{G_{n}^2}{\kappa})$ ($n=1,2$). Moreover, the coupling pulse is much stronger than the probe pulse so that the solution to Eq. (\ref{3}) can take the form $\langle o\rangle=o_{0}+o_{+}e^{-i\delta t}+o_{-}e^{i\delta t}$ ($o=Q_{n}$, $P_{n}$, or $c$), then a set of coupled differential equations $o_{0}$, $o_{+}$, and $o_{-}$ can be derived from Eq. (\ref{3}), and they can be solved numerically.

\begin{figure}[!h]
\begin{center}
\scalebox{0.6}{\includegraphics{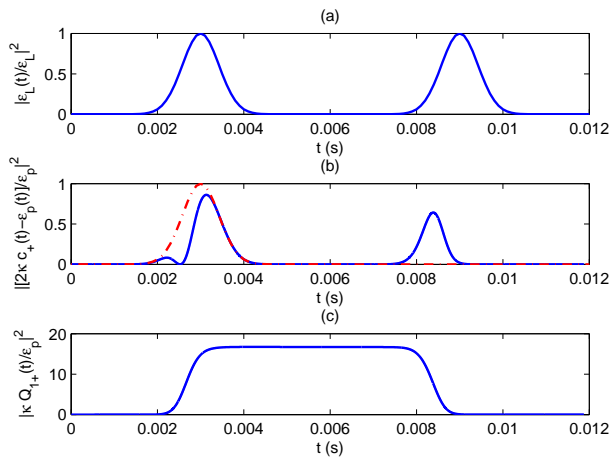}} \caption{\label{Fig8} (Color online) (a) The normalized power $|\varepsilon_{L}(t)/\varepsilon_{L}|^2$ of the applied coupling field as a function of time $t$ (s). (b) The normalized power $|\varepsilon_{p}(t)/\varepsilon_{p}|^2$ of the input probe field (dot-dashed curve) and the normalized output probe power $|[2\kappa c_{+}(t)-\varepsilon_{p}(t)]/\varepsilon_{p}|^2$ (solid curve) as a function of time $t$ (s). (c) The normalized intensity $|\kappa Q_{1+}/\varepsilon_{p}|^2$ of the mechanical excitation of the membrane at frequency $\omega_{1}$ as a function of time $t$ (s). Parameter: $\delta=\omega_{1}$.}
\end{center}
\end{figure}

\begin{figure}[!h]
\begin{center}
\scalebox{0.6}{\includegraphics{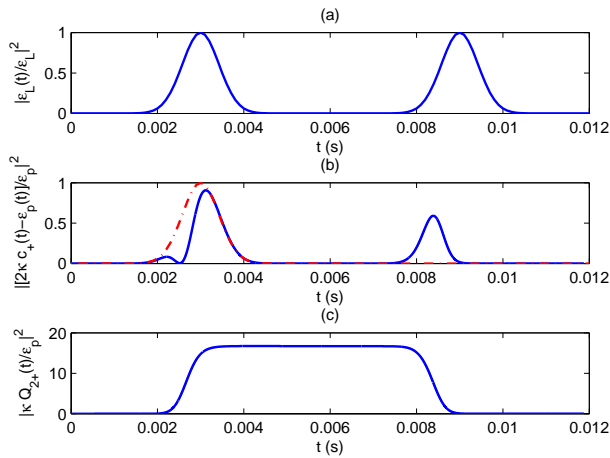}} \caption{\label{Fig9} (Color online) (a) The normalized power $|\varepsilon_{L}(t)/\varepsilon_{L}|^2$ of the applied coupling field as a function of time $t$ (s). (b) The normalized power $|\varepsilon_{p}(t)/\varepsilon_{p}|^2$ of the input probe field (dot-dashed curve) and the normalized output probe power $|[2\kappa c_{+}(t)-\varepsilon_{p}(t)]/\varepsilon_{p}|^2$ (solid curve) as a function of time $t$ (s). (c) The normalized intensity $|\kappa Q_{2+}/\varepsilon_{p}|^2$ of the mechanical excitation of the membrane at frequency $\omega_{2}$ as a function of time $t$ (s). Parameter: $\delta=\omega_{2}$.}
\end{center}
\end{figure}

We use the following parameters for numerical simulations: $\omega_{m}=2\pi\times134$ kHz, $\omega_{1}=\omega_{m}+0.05\omega_{m}$, $\omega_{2}=\omega_{m}-0.05\omega_{m}$, $\kappa=\omega_{m}/5$, $\gamma_{1}=\gamma_{2}=2\pi\times 0.12$ Hz, $g_{1}=g_{2}=0.0008\kappa$, $\lambda=1064$ nm, $\wp=0.04$ $\mu$W, $\Delta=\omega_{m}$, $\tau_{p}=\tau_{L}=0.6$ ms, $t_{wr}=3$ ms, $t_{rd}=9$ ms. Hence $\tau_{p}^{-1}=2\pi\times265$ Hz, which is less than the FWHM of EIT dip ($\gamma_{1}+\frac{G_{1}^2}{\kappa}=\gamma_{2}+\frac{G_{2}^2}{\kappa}=2\pi\times1678$ Hz),

We plot the normalized power $|\varepsilon_{L}(t)/\varepsilon_{L}|^2$ of the input coupling field, the normalized power $|\varepsilon_{p}(t)/\varepsilon_{p}|^2$ of the input probe field, the normalized output probe power $|[2\kappa c_{+}(t)-\varepsilon_{p}(t)]/\varepsilon_{p}|^2$, and the normalized intensity of the mechanical excitation $|\kappa Q_{j+}/\varepsilon_{p}|^2$ for $\delta=\omega_{n}$ ($n=1,2$) in Figs.~\ref{Fig8} and \ref{Fig9}. For $\delta=\omega_{1}$, when a writing control pulse is applied, the EIT occurs due to the nonlinear interaction between the input coupling pulse and the movable membrane at frequency $\omega_{1}$, it is seen that a fraction of the probe pulse leaves the cavity, leading to the first peak of the solid curve in Fig.~\ref{Fig8}(b), while a part of the probe pulse is converted into the mechanical excitation of the membrane at frequency $\omega_{1}$ via the downconversion process $\omega_{p}-\omega_{c}=\omega_{1}$ and stored for a storage time of about $t_{rd}-t_{wr}=6$ ms (Fig.~\ref{Fig8}(c)), and then retrieved by applying a reading coupling pulse via the upconversion process $\omega_{c}+\omega_{1}=\omega_{p}$ (the second peak of the solid curve in Fig.~\ref{Fig8}(b)). For $\delta=\omega_{2}$, the solid curves in Fig.~\ref{Fig9}(b)(c) show a similar result. Therefore, a probe pulse at two different frequencies $\omega_{p}=\omega_{c}+\omega_{1}$ and $\omega_{p}=\omega_{c}+\omega_{2}$ can be stored and retrieved by an optical cavity containing two membranes.


\section{Conclusions}

In conclusion, we have discussed the response of an optomechanical system which includes $N$ membranes having slightly different frequencies to a weak probe field in the presence of an intense coupling field. We have given a general analytical expression for the output probe field. We have shown that the system can exhibit the phenomenon of multiple EIT. At most $N$ transparency windows can be observed. The linewidths of the transparency windows increase with the effective optomechanical strengths. The
 dispersion behavior of the probe field shows steep anomalous dispersion at the EIT dips, which results in negative group velocities for the output probe field at multiple
frequencies. At most $N$ different negative values of the group velocities at the transparency can be obtained. These group velocities can be controlled by altering the effective optomechanical strengths. Such controllable multiple EIT windows in a macroscopic optomechanical system with $N$ membranes permit the probe field at different frequencies transparent simultaneously, hence this system has potential applications in multi-channel optical communication and multi-channel quantum information processing. Further this optomechanical design can be employed to simultaneously realize optical memories for the probe field at different frequencies, leading to higher efficiencies in optical communication.

\section*{ACKNOWLEDGMENTS}
This work was supported by the Singapore National Research Foundation under NRF Grant No. NRF-NRFF2011-07.

\end{document}